\documentstyle[prl,aps,graphicx,twocolumn]{revtex}

\def\be{\begin{equation}}
\def\ee{\end{equation}}
\def\ba{\begin{eqnarray}}
\def\ea{\end{eqnarray}}

\begin{document}
\draft
\title{Two ground-state modifications of quantum-dot beryllium
}

\author{S. A. Mikhailov\footnote{Present address: Max-Planck Institut f\"ur Festk\"orperforschung, Heisenbergstr. 1, 70569 Stuttgart, Germany. Electronic address: S.Mikhailov@fkf.mpg.de}}
\address{Theoretical Physics II, Institute for Physics, University of Augsburg, 86135 Augsburg, Germany}

\date{\today}
\maketitle
\begin{abstract}
Exact electronic properties of a system of four Coulomb-interacting two-dimensional electrons in a parabolic confinement 
are reported. We show that degenerate ground states of this system are characterized by qualitatively different internal electron-electron correlations, and that the formation of Wigner molecule in the strong-interaction regime is going on in essentially different ways in these ground states.
\end{abstract}

%73.21.La Quantum dots
%71.10.-w Theories and models of many-electron systems
%73.20.Qt Electron solids

Quantum dots \cite{Jacak97}, -- artificial electron systems realizable in modern semiconductor structures, -- are ideal physical objects for studying effects of electron-electron correlations. They contain a few ($N$) two-dimensional (2D) interacting electrons moving in the plane $z=0$ in a lateral confinement potential $V(x,y)$. The potential $V$ can often be modeled by a harmonic oscillator form $V=K(x^2+y^2)/2$, with a characteristic frequency $\omega_0=\sqrt{K/m^\star}$ and a characteristic %oscillator 
length $l_0=\sqrt{\hbar/m^\star\omega_0}$ (here $m^\star$ is the effective mass of electrons). The relative strength of interaction and kinetic effects can be characterized by the dimensionless parameter 
\begin{equation}
\lambda=\frac{l_0}{a_B}=\sqrt{\frac{e^2/a_B}{\hbar\omega_0}}\propto \frac{e^2}{\hbar^{3/2}},
\end{equation} 
determined by the ratio of quantum-mechanical and interaction length (energy) scales ($a_B$ is the effective Bohr radius of the host semiconductor). The curvature of the confinement $K$, and hence the interaction parameter $\lambda$, can be experimentally varied, so that quantum dots are often referred to as artificial atoms with tunable physical properties. Theoretical study of physical properties of such ``atoms'' {\em as a function of} the Coulomb-interaction parameter $\lambda$ may give a valuable information about the role of electron-electron correlations in interacting quantum systems. The physics of interactions becomes especially interesting in zero external magnetic field $B$, when electronic spins are not polarized and are active players in the game. Detailed theoretical study of physical properties of quantum-dot atoms, including the Fermi-liquid -- Wigner-molecule crossover in the ground state with growing strength of intra-dot Coulomb interaction, is one of the challenging problems in this field, which was attracting increasing interest in recent years\cite{Egger99,Creffield99,Yannouleas00,Hausler00,Reimann00,Pederiva00,Filinov01,Matulis01,Reusch01,Akbar01,Varga01,Mikhailov02a,Mikhailov02b,Harju02} (for more references see \cite{Mikhailov02b} and a recent review \cite{Maksym00}). 

The Hamiltonian of an $N$-electron dot
\begin{equation}
\hat H =\sum_{i=1}^N \left(\frac {\hat {\bf p}_i^2}{2m^\star} +
\frac{m^\star\omega_0^2 {\bf r}_i^2}2\right)
+ 
\frac 12\sum_{i\neq j=1}^N \frac{e^2}{%\kappa
|{\bf r}_i-{\bf r}_j|} ,
\label{qdhamiltonian}
\end{equation}
commutes with the operators of the total angular momentum $L\equiv L_z$, total spin $S$, and its projection $S_\zeta$ on some axis. Solutions 
\be
E_{nl}=\hbar\omega_0(2n+|l|+1)
\ee
of the single-particle problem of quantum-dot {\em hydrogen} ($N=1$) \cite{Fock28} are characterized by the radial $n$ and asimutal $l$ (angular momentum) quantum numbers. The ground state of quantum-dot {\em helium} ($N=2$) \cite{Merkt91} is characterized by the total angular momentum $L=0$ and the total spin $S=0$. In the limit of weak interaction $\lambda\ll 1$ two electrons with opposite spins occupy the single-particle state $(n,l)=(0,0)$, forming the configuration $(0,0,\uparrow\downarrow)$. The state $(L,S)=(0,0)$ remains the ground state at any $\lambda$, also in the regime of strong interaction. The first excited state of quantum-dot helium is the state $(L,S)=(1,1)$.  

In quantum-dot {\em lithium} $(N=3)$ \cite{Mikhailov02b} the $\lambda$-dependence of the ground-state properties is more complicated. In the regime of weak interaction $\lambda\ll 1$ electrons form the configuration $[(0,0,\uparrow\downarrow),(0,1,\uparrow)]$ with $(L,S)=(1,1/2)$. The state $(1,1/2)$ remains the ground state up to $\lambda=\lambda_c\approx 4.343$, where a transition to another ground state with $(L,S)=(0,3/2)$ occurs. All physical properties of the dot sharply change at the transition point. At $\lambda>\lambda_c$ electron distribution in the dot has the form of an angle-averaged equilateral triangle (the Wigner molecule). 

The goal of this Letter is to present {\em exact} results of solution of the four-electron quantum-dot {\em beryllium} problem at $B=0$. Some results for the energy of states at a few points on the $\lambda$-axis, obtained by different variational methods, were reported in the literature (e.g. \cite{Egger99,Pederiva00,Varga01}). We complete and essentially improve these results, presenting them for the ground and a number of excited states in a broad range and {\em as a function of} $\lambda$. A principally new finding of this work results from a detailed study of ground-state properties of quantum-dot beryllium. The ground state of this atom is degenerate with respect to the total spin projection $S_\zeta$, and we show that electron-electron {\em correlation functions}, as well as their {\em spin density}, are substantially different in different ground states. As a result, formation of Wigner molecules with the growing interaction strength is going on in qualitatively different ways, dependent on the value of the total spin projection $S_\zeta$ ($0$ or $\pm 1$). This result is valid not only in the considered parabolic quantum-dot model. It has a general character and should be the case in any strongly-interacting many-electron system (e.g. in natural atoms) with degenerate ground states. 

The Schr\"odinger equation with the Hamiltonian (\ref{qdhamiltonian}) and $N=4$ is solved by the exact-diagonalization method described in Ref.\cite{Mikhailov02b}. The many-body wave function is expanded in a complete set $\Psi_u$ of many-body eigenstates of the Hamiltonian $\hat H_0$ of non-interacting particles, $\hat H_0\Psi_u=E_u\Psi_u$, with the number of basis states $N_{mbs}$ restricted by the condition $E_u\le E_{th}$, where $E_{th}$ is a threshold value. The larger the threshold energy $E_{th}$, the higher the accuracy of the calculated energy at a given value of $\lambda$, and the broader the range of $\lambda$'s, in which results have a given accuracy. We performed calculations until a sufficient (very high) accuracy has been achieved in the whole (studied) range of $\lambda$ ($\lambda\le 10$). For the ground state energy our accuracy is about two orders of magnitude higher than in corresponding Quantum Monte Carlo (QMC) calculations; at $\lambda=10$ it comprises about $6\times 10^{-4}$\% (see Table II below).

\begin{figure}
\includegraphics[width=8.2cm]{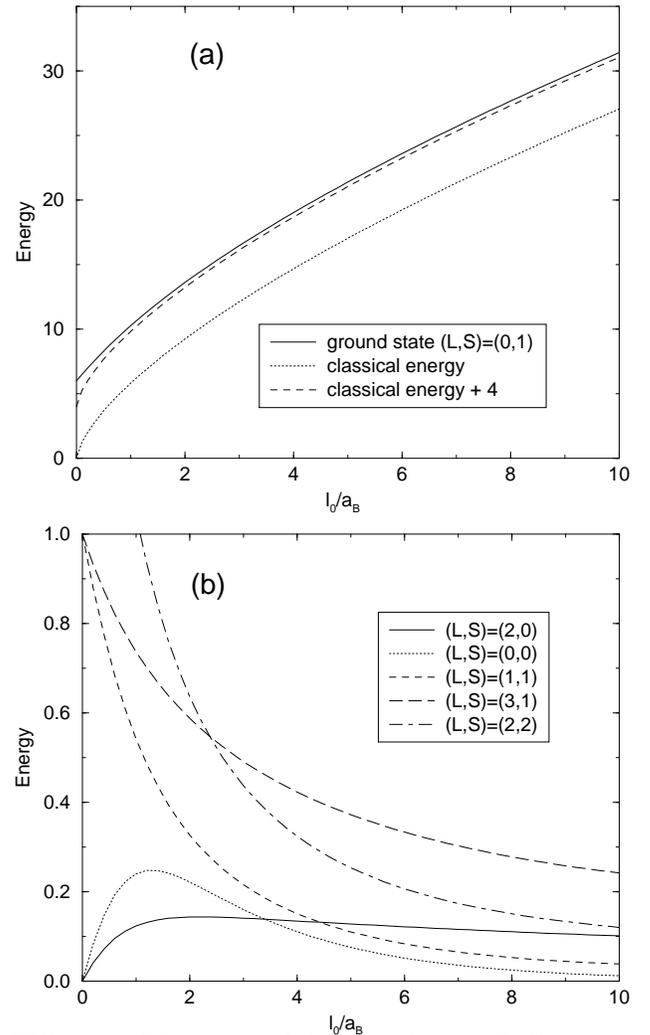}
\caption{(a) The energy of the ground state $(L,S)=(0,1)$ of quantum-dot beryllium as a function of the interaction parameter $\lambda$, together with the clasical contribution $E_{cl}$ and the energy $E_{cl}+4\hbar\omega_0$. (b) The energies of a few excited states $E_{(LS)}-E_{(01)}$ (the ground state energy subtracted) as a function of $\lambda$. Calculations were performed with $E_{th}=22\hbar\omega_0$, $N_{mbs}=14453$. The energy unit on both plots is $\hbar\omega_0$. 
}
\label{groundstate}
\end{figure}

 Figure \ref{groundstate}a shows the ground-state energy of a four-electron parabolic quantum dot as a function of the interaction parameter $\lambda$, together with the classical contribution 
\begin{equation}
E_{cl} = 6 [(1+\sqrt{8})/4]^{2/3}(e^2/l_{cl}),\ \ l_{cl}=(e^2/m^\star\omega_0^2)^{1/3},
\label{infinitelmbda}
\end{equation}
(the energy of four Coulomb-interacting point-like particles in a parabolic potential \cite{Bolton93,Bedanov94}), and with the curve $E_{cl}+4\hbar\omega_0$ (the classical energy plus $\hbar\omega_0$ per particle). The ground state is characterized by quantum numbers $(L,S)=(0,1)$ and is three-fold degenerate. In the weak-interaction regime the corresponding wave functions are $\Psi_1$, $(\Psi_2+\Psi_3)/\sqrt{2}$ and $\Psi_4$ ($S_\zeta=+1,0,$ and $-1$, respectively), where 
\ba
\Psi_1&=&[(0,0,\uparrow\downarrow),(0,-1,\uparrow),(0,1,\uparrow)], \nonumber \\
\Psi_2&=&[(0,0,\uparrow\downarrow)
,(0,-1,\uparrow),(0,1,\downarrow)], \nonumber \\
\Psi_3&=&[(0,0,\uparrow\downarrow)
,(0,-1,\downarrow),(0,1,\uparrow)], \nonumber \\
\Psi_4&=&[(0,0,\uparrow\downarrow)
,(0,-1,\downarrow),(0,1,\downarrow)].
%\label{smalllambda}
\ea
Figure \ref{groundstate}b shows the energies of a few excited states (the ground-state energy is  subtracted). At small $\lambda$ the first and the second excited states are $(L,S)=(2,0)$ and $(L,S)=(0,0)$ respectively. The state $(L,S)=(2,0)$ is two-fold degenerate and corresponds, in the limit $\lambda\to 0$, to the configurations $[(0,0,\uparrow\downarrow),(0,-1,\uparrow\downarrow)]$ and $[(0,0,\uparrow\downarrow),(0,1,\uparrow\downarrow)]$. The state $(L,S)=(0,0)$ is non-degenerate and corresponds to the configuration $(\Psi_2-\Psi_3)/\sqrt{2}$ at $\lambda\to 0$. These and some other excited states are listed in Table \ref{tab1} in the order of increase of their energies at $\lambda\ll 1$. At larger $\lambda$'s the order of excited levels changes, as seen from Figure \ref{groundstate}b. The level crossings are the case at $\lambda\approx 3.40$ [the states $(L,S)=(0,0)$ and $(2,0)$], $\lambda\approx 4.42$ [the states $(1,1)$ and $(2,0)$], and $\lambda\approx 2.37$ [the states $(2,2)$ and $(3,1)$]. The state $(L,S)=(0,1)$ remains the ground state at all (studied) values of $\lambda$ ($\le 10$).

Table \ref{tab2} exhibits numerical data, at a few $\lambda$ points, for the ground state $(L,S)=(0,1)$, and for one of the excited (fully spin-polarized) state $(L,S)=(2,2)$. Here we also compare our results with QMC calculations from Ref. \cite{Egger99}. Notice that, although the QMC results are, in general, in a good agreement with exact ones for the fully spin-polarized (excited) state, they have an essential systematic error for the ground state $(L,S)=(0,1)$: for instance, at $\lambda=10$ the absolute error of the QMC method (about $0.07\hbar\omega_0$) is substantially larger than the energy difference between the states $(0,1)$ and $(0,0)$ (estimated as $<0.013\hbar\omega_0$ at $\lambda=10$). 

\begin{figure}
\includegraphics[width=8.2cm]{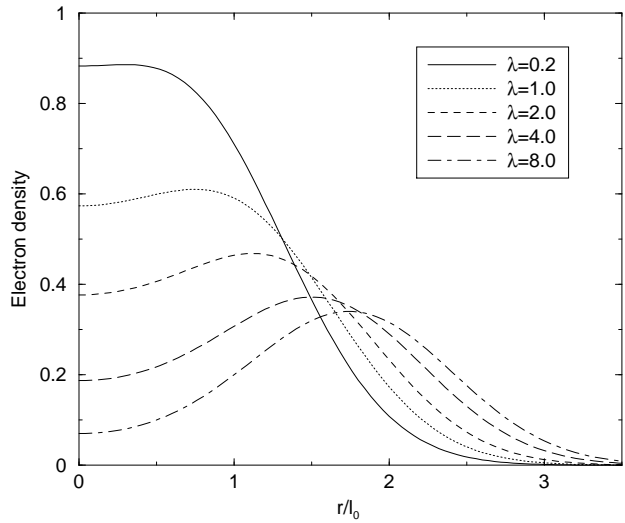}
\caption{Electron density $\pi l_0^2n_{\sigma}(r)$ of spin-up and spin-down polarized electrons [$n_\uparrow(r)\equiv n_\downarrow(r)$] as a function of radial coordinate $r/l_0$ in the ground state $(L,S)=(0,1)$ with $S_\zeta=0$ at a few values of the interaction parameter $\lambda$.
}
\label{densityplot0}
\end{figure}

\begin{figure}
\includegraphics[width=8.2cm]{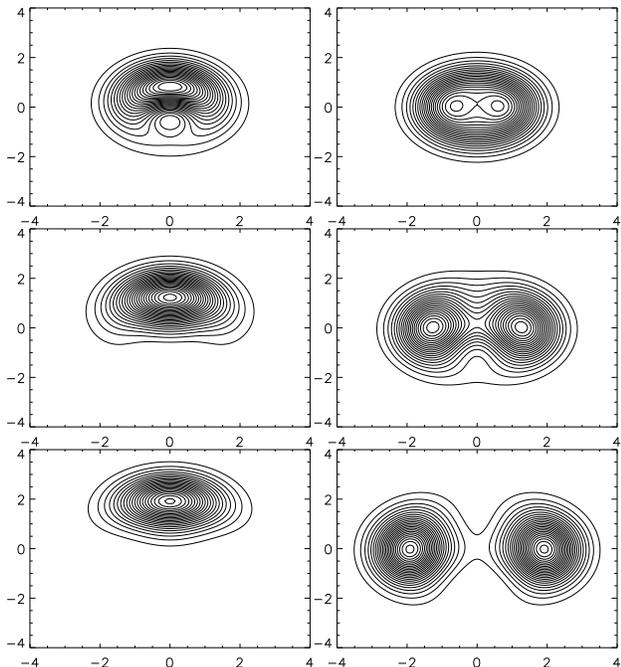}
\caption{Pair-correlation functions $P_{\uparrow\uparrow}({\bf r},{\bf r}^\prime)\equiv P_{\downarrow\downarrow}({\bf r},{\bf r}^\prime)$ (left column) and $P_{\downarrow\uparrow}({\bf r},{\bf r}^\prime)\equiv P_{\uparrow\downarrow}({\bf r},{\bf r}^\prime)$ (right column) in the ground state $(L,S)=(0,1)$ with $S_\zeta=0$, as a function of ${\bf r}$ at ${\bf r}^\prime=(0,-R_{cl})$ (the length unit is $l_0$). The interaction parameter $\lambda$ assumes the values $\lambda=0.2$, 2, and 8, from up to down ($R_{cl}/l_0=0.57$, 1.24, and 1.97, respectively).
}
\label{pcf0}
\end{figure}

\begin{figure}
\includegraphics[width=8.2cm]{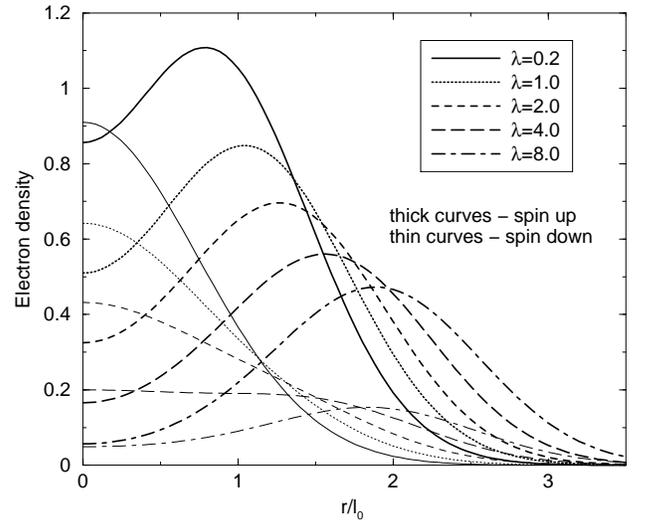}
\caption{The same as in Figure \protect\ref{densityplot0}, but for the ground state with $S_\zeta=+1$. The total (charge) density $n_\uparrow(r)+n_\downarrow(r)$ is the same as in Figure \protect\ref{densityplot0}.
}
\label{densityplot1}
\end{figure}

\begin{figure}
\includegraphics[width=8.2cm]{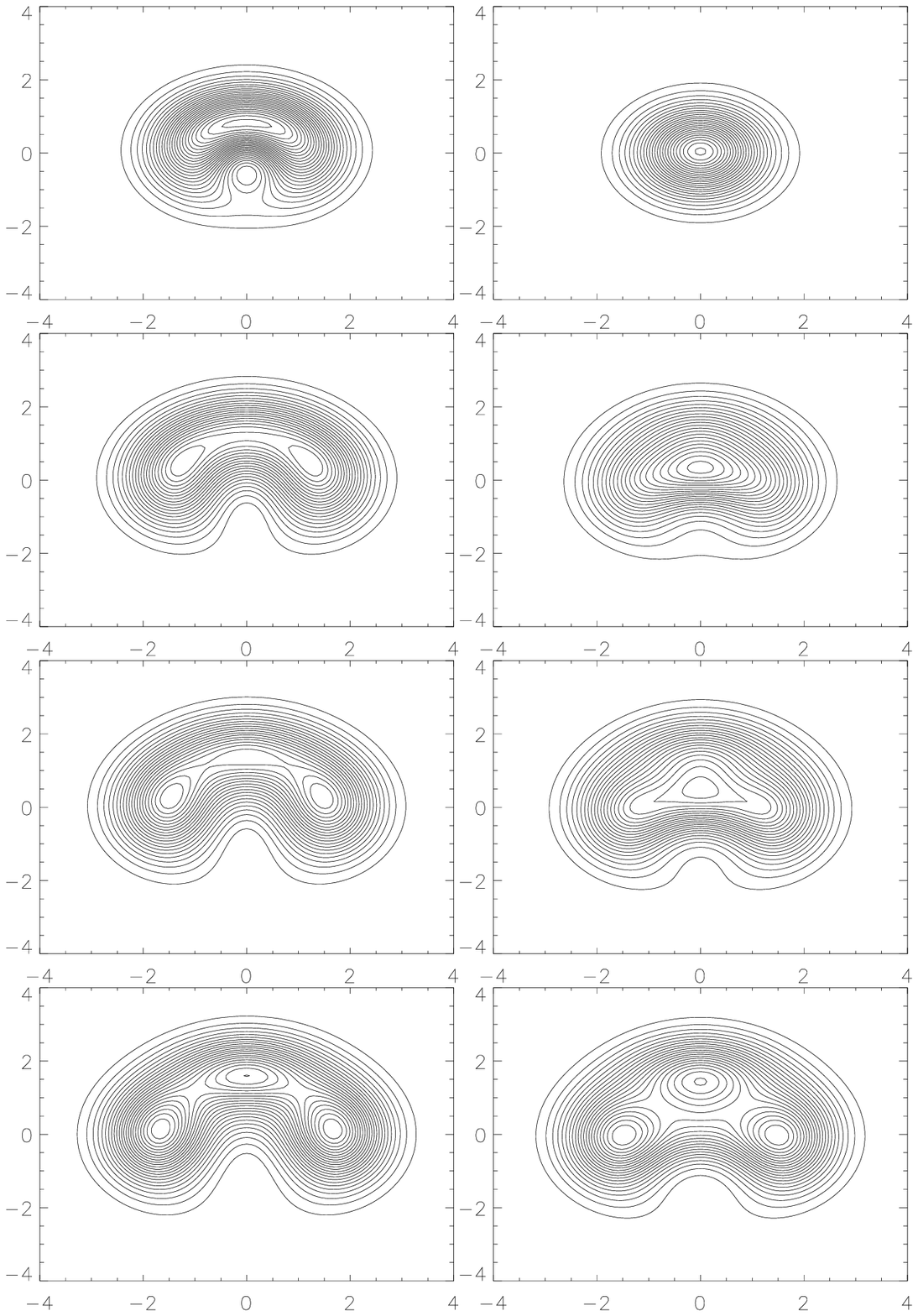}
\caption{Pair-correlation functions $P_{\uparrow\uparrow}({\bf r},{\bf r}^\prime)$ (left column), and $P_{\downarrow\uparrow}({\bf r},{\bf r}^\prime)$  (right column) in the ground state $(L,S)=(0,1)$ with $S_\zeta=+1$, as a function of ${\bf r}$ at ${\bf r}^\prime=(0,-R_{cl})$ (the length unit is $l_0$, the second subscript corresponds to the electron fixed at the ${\bf r}^\prime$ point). The interaction parameter $\lambda$ assumes the values $\lambda=0.2$, 2, 3, and 4.5, from up to down ($R_{cl}/l_0=0.57$, 1.24, 1.42, and 1.62, respectively).
}
\label{pcf1}
\end{figure}

Now consider physical properties of quantum-dot beryllium in the ground states $(L,S,S_\zeta)=(0,1,0)$ and $(L,S,S_\zeta)=(0,1,+1)$. These states are essentially different. In the ground state with $S_\zeta=0$ two electrons are polarized up and two electrons are polarized down. They form a symmetric square structure, with two up- and two down-polarized electrons occupying opposite corners of a square, see Figures \ref{densityplot0} and \ref{pcf0} showing the density and the pair-correlation functions of electrons in this state. The relative electronic distribution holds its square form at all $\lambda$'s, also in the regime of small interactions $\lambda\ll 1$. In the ground state with $S_\zeta=+1$ three electrons are polarized up and one electron is polarized down. In the weak- and moderate-interaction regimes ($\lambda\lesssim 2$) they form a {\em triangular} structure with one (down) electron occupying the center of the dot, and three (up) electrons rotating around, see Figures \ref{densityplot1} and \ref{pcf1}. When $\lambda$ increases, the growing Coulomb interaction pushes the down-electron out from the center, and the triangular configuration is smoothly transformed to a (classical-type) square-form structure. In Figure \ref{pcf1} one sees how this process begins at $\lambda\simeq 3$ (the third row of plots) and how it is completed at $\lambda\simeq 4.5$ (the last row). Thus, correlations between electrons with different spins are essentially different in ground states with $S_\zeta=0$ and $S_\zeta=\pm 1$, and the formation of Wigner molecules is going on in qualitatively different ways in these states.

The coexistence of degenerate ground states with qualitatively different electron-electron correlations is obviously independent of the 2D nature of the electrons and of the parabolic form of the confinement potential in the present model. The same situation should be also the case, for example, in a natural carbon atom, which has $S=1$ in the ground state. In a hypothetic case of a quantum system with $S=2$ and more particles one should also expect that degenerate ground states with $S_{\zeta}=0$, $\pm 1$ and $\pm 2$ are characterized, in the strong-interaction regime, by qualitatively different inter-electron correlations. 

To conclude, we have studied energy spectra, charge and spin densities, and electron-electron correlations in quantum-dot beryllium -- a system of four strongly interacting electrons in a harmonic oscillator potential. It was shown that this artificial atom can exist in two essentially different ground states, characterized by qualitatively different electron-electron correlation functions. 

%\acknowledgments
The work was supported by the Deutsche Forschungsgemeinschaft through SFB 484.

%\bibliography{$HOME/BIB-FILES/dots,$HOME/BIB-FILES/fqhe,$HOME/BIB-FILES/mikhailov,$HOME/BIB-FILES/wc}
%\bibliographystyle{/usr/local/server/teTeX/local/tex/latex/contrib/other/revtex/prsty}

\begin{table}
\caption{The ground and a few lowest excited states of quantum-dot beryllium: the total angular momentum and spin $(L,S)$, energy in the weak-interaction limit $E_{LS}(\lambda\to 0)$ (in units $\hbar\omega_0$), and degeneracy $g$ of the states. The states are shown in the order of increasing of their energy at $\lambda\ll 1$. 
\label{tab1}
}
\begin{tabular}{cccc}
No&$(L,S)$ & $E_{LS}(\lambda\to 0)$ & $g$\\
\tableline
1&$(0,1)$ & 6 & 3 \\
2&$(2,0)$ & 6 & 2 \\
3&$(0,0)$ & 6 & 1 \\
4&$(1,1)$ & 7 & 6 \\
5&$(3,1)$ & 7 & 6 \\
6&$(1,0)$ & 7 & 2 \\
7&$(1,1)$ & 7 & 6 \\
8&$(1,1)$ & 7 & 6 \\
9&$(1,0)$ \& $(3,0)$ & 7 & 4 \\
10&$(1,0)$ & 7 & 2 \\
11&$(2,2)$ & 8 & 10 \\
12&$(0,2)$ & 8 & 5 \\
\end{tabular}
\end{table}

\begin{table}
\caption{Energies of the ground state $(L,S)=(0,1)$ and of the fully spin-polarized state $(L,S)=(2,2)$ in quantum-dot beryllium from exact diagonalization (this work) and Quantum Monte Carlo calculations \protect\cite{Egger99}. Exact data in the first five lines were obtained with $E_{th}=22\hbar\omega_0$, which corresponds to $N_{mbs}=14453$ for the ground state and $N_{mbs}=3248$ for the spin polarized state. The data in the three last lines were obtained with $E_{th}=24\hbar\omega_0$ ($N_{mbs}=24348$) for the state $(0,1)$ and with $E_{th}=26\hbar\omega_0$, $N_{mbs}=8721$ for the state $(2,2)$.
\label{tab2}
}
\begin{tabular}{rcccc}
$\lambda$ & (0,1) & (0,1)\tablenotemark[1] & (2,2) & (2,2)\tablenotemark[1] \\
\tableline
2 & 13.6180 & 13.78(6) & 14.2535 & 14.30(5) \\
4 & 19.0323 & 19.15(4) & 19.3565 & 19.42(1) \\
6 & 23.5958 & 23.62(2) & 23.8025 & 23.790(12) \\
8 & 27.6696 & 27.72(1) & 27.8203 & 27.823(11) \\
10& 31.4122 & 31.48(2) & 31.5326 & 31.538(12) \\
\tableline
10& 31.4120 & 31.48(2) & 31.5323 & 31.538(12) \\
15& 39.8163 &          & 39.8970 & \\
20& 47.4002 &          & 47.4013 & \\
\end{tabular}
\tablenotetext[1]{Ref. \protect\cite{Egger99}}
\end{table}

\end{document}